\documentstyle[aps,epsf,prl,twocolumn]{revtex} 
\def\B.#1{{\bbox{#1}}}

\newcommand{\eq}[1]{\begin{equation}\label{#1}}
\newcommand{\eqs}{\begin{equation}}
\newcommand{\en}{\end{equation}}
\newcommand{\bea}{\begin{eqnarray}}
\newcommand{\ena}{\end{eqnarray}}
\newcommand{\r}{{\bf r}}
\newcommand{\Z}{{\bf Z}}

\newcommand{\h}{{\bf h}}
\newcommand{\U}{{\bf u}}

\begin{document}
\twocolumn[\hsize\textwidth\columnwidth\hsize\csname@twocolumnfalse\endcsname
\title{ Anomalous scaling in passive scalar advection: Monte-Carlo
Lagrangian trajectories}
\author{
Omri Gat(1)(2), Itamar Procaccia(1) and Reuven Zeitak(1)(3)\\
(1)Dept. Chem. Physics, The Weizmann Institute of Science,
Rehovot 76100, Israel\\
(2) Dept. Physique Theorique, Universite de Geneve, 32, Bld d'Yvoy, 1211
Geneve 4, Switzerland\\
(3) Laboratoire de Physique Statistique, ENS,
24 rue Lhomond, 75231 Paris Cedex 05, France\\ }
\maketitle
\widetext
\begin{abstract}
We present a numerical method which is used to calculate anomalous scaling
exponents
of structure functions in the Kraichnan passive scalar advection model
(R. H. Kraichnan, Phys. Fluids {\bf11}, 945 (1968)). This Monte-Carlo method,
which is applicable in any space dimension, is based on the
Lagrangian path interpretation of passive scalar dynamics, and uses the
recently
discovered equivalence between scaling exponents of structure functions and
relaxation rates in the
stochastic shape dynamics of groups of Lagrangian particles. We calculate third
and fourth order anomalous exponents for several dimensions, comparing with the
predictions of perturbative
calculations in large dimensions. We find that Kraichnan's closure appears
to give
results in close agreement with the numerics. The third order
exponents are compatible
with our own previous nonperturbative calculations.
\end{abstract}
\vskip 5mm
]
\narrowtext
The Kraichnan rapid advection model \cite{kr68} is a simplified model for
turbulent
advection of a passive scalar $T(\B.r)$ in which the velocity field scales
in space but
decorrelates in time infinitely rapidly. The scaling properties
of this model are interesting:
The structure functions $S_n(R)\equiv \langle
[T(\B.r+\B.R)-T(r)]^n\rangle$ depend on $R$ like power laws, $S_n(R)\sim
R^{\zeta_n}$,
and the exponents $\zeta_n$ are ``multiscaling" in the sense that they are
nonlinear
functions of the index $n$ in contradiction with classical (Kolmogorov)
approaches. The
importance of this model lies in the fact that it has been the first
example in which the
mechanism for multiscaling, which appears also in Navier-Stokes turbulence
\cite{legacy},
has been identified\cite{kr94}, and in which a controlled calculation of
scaling
exponents is possible. As such it is an important laboratory for testing
ideas that
may pertain also for the understanding of multiscaling in Navier-Stokes
turbulence.
It is not surprising therefor that it drew enormous attention and
theoretical efforts.
To date, the scaling exponents in the Kraichnan model
have been calculated only
in a small part of the parameter space, which consists of the velocity scaling
exponent~$0\le\xi\le2$ (see below), and the space dimensionality~$d$.
Perturbative calculations were performed in the nearly integrable
limits $\xi\ll1$ \cite{gk}, $2-\xi\ll1$ \cite{ssp,3pt-2,balkovsky}, and $d\gg1$
\cite{grisha,shape}. In addition, Kraichnan proposed \cite{kr94} an interesting
closure method to estimate the exponents for even
$n$ and arbitrary $d$. Direct
numerical simulations were performed only in two dimensions \cite{kyc,barak}.
The exponents $\zeta_2$ and $\zeta_3$ were computed throughout the
parameter space,
with an explicit solution in the former case, and by a finite difference
scheme in
the latter \cite{3pt-1}.

In a recent article \cite{shape}, we presented a new interpretation of the
phenomenon of
anomalous scaling in passive scalar correlation functions: the theory
connects between
the scaling exponents $\zeta_n$ and the
relaxation rates of the geometric shape formed by $n$ points advected
by the turbulent flow.
Starting from a generic distribution in the space of shapes,
the
stochastic advection can be considered as a relaxation to the invariant
measure in
this space, while the over-all scale undergoes a Richardson diffusion. In
\cite{shape} this identification was used to perform $1/d$ perturbative
calculations of the scaling exponents. In this Letter we apply the theory,
using
Monte-Carlo methods, to calculate the scaling exponent throughout the entire
parameter space\cite{massimo}. We generate random sample paths for the $n$
trajectories,
and measure the shape relaxation rate by the standard method of analyzing the
autocorrelation of the signal. Compared to direct numerical simulation,
this method
demands relatively modest computing resources:
there is no need to keep track of the whole scalar field. The realizations of
random velocity are generated only at the instantaneous
positions of the advected points,
and this amounts to sampling a small set of correlated Gaussian random
variables.
Hence, it is possible to perform simulations in arbitrary integer
dimensions. Another
advantage of the Monte-Carlo method is that it is trivially parallelizable,
in the
sense that the data is collected from numerous independent realizations, whose
calculation is completely independent, and may therefore be performed on
different
machines. Nevertheless, the accurate calculation of the anomalous exponents
is still
not easy, due to slow convergence rate and the large number of realizations
that is
therefore required.

The Kraichnan model describes the advection of a passive scalar
by a random velocity
field $\U$ which is Gaussian, $\delta$-correlated in time, incompressible and
self similar in space:
\bea
\langle(\U(\r,t)-\U(\r',t))&\otimes&(\U(\r,t')-\U(\r',t')) \rangle
\nonumber\\ &=&\h(\r-\r')\delta(t-t')\ , \nonumber\\
\h(\r)&=&\left({r\over \ell}\right)^\xi
({\bf1}-{\xi\over d-1+\xi}{\r\otimes\r\over r^2}).
\ena
The passive scalar is driven by a large scale forcing (of scale $L$),
which is taken Gaussian and $\delta$-correlated in time as well.
The fundamental objects of the Kraichnan model are the many point
correlation functions
\begin{equation}
{\cal F}_{2n}({\bf r}_1,{\bf r}_2,...,{\bf r}_{2n})
\equiv \langle T(\B.r_1,t)
T(\B.r_2,t)\dots T(\B.r_{2n},t)\rangle \ , \end{equation}
where pointed brackets denote an ensemble average with respect to the
stationary statistics of the forcing {\em and } the statistics of the
velocity field.

The correlation functions may be expanded in a series of scale
invariant terms, {\it i.e.}, terms which are homogeneous functions of their
variables.
The most important term for $R/L\to 0$ is expected to be a zero mode of the
Kraichnan
partial differential operator operator\cite{kr94,grisha}.
The functions ${\cal F}_{2n}$ contain, in addition
to their own zero-modes, also contributions from the zero modes of lower
order functions and normal scaling contributions due to the external
forcing. These do not contribute however to the $n$th order structure
functions, and thus do not affect the scaling exponent $\zeta_n$. The question
is how to compute the homogeneity exponents of the zero modes. Kraichnan
himself
avoided this (difficult) problem and suggested a closure procedure whose main
result is a closed form ansatz for the scaling exponents
\begin{equation}
\zeta_{2n}={1\over
2}\left[\zeta_2-d+\sqrt{(\zeta_2+d)^2+4\zeta_2(n-1)}\right] \
, \label{kra}
\end{equation}
where $\zeta_2=2-\xi$.
We will compare the $\xi=1$ line of this prediction to our numerics.

We explain now shortly how multiscaling of scalar correlation functions
can be obtained from the random evolution of shapes as a function of the
overall scale. More details can be found in \cite{shape}. We consider $n$
points advected simultaneously by the random velocity field $\U$, \bea
\partial_{t} \r_i(t)&=&\U(\r_i(t),t),\qquad 1\le i\le n.\label{traject1}
\ena

The configuration of these $n$ points can be described at each instant by one
overall scale $s$, and a set of dimensionless variables $\Z$ which describe
its geometry. For example, in the case $n=3$, the overall scale could be
defined
by the sum of edge lengths, and the shape by two of the angles of the triangle
defined by the three points. As far as scaling properties are considered the
precise definition of the overall scale and the shape variables is unimportant,
and see \cite{shape} for details.

The evolution of the shape defined by our $n$ advected particles is
characterized
by an operator that we denote by $\gamma_n(\Z_{0},\Z,{s\over s_{0}})$.
This operator determines the probability that, starting at scale $s_0$ with
shape
$\Z_0$, the $n$ points will form the shape $\Z$, for the first time that the
overall scale reaches the value $s>s_0$. We are interested in the asymptotics
of
the shape evolution operator; its large $s$ expansion is
\eq{gamma}
\gamma_n(\Z_0,\Z,{s\over s_0})=\sum_{m=0}^\infty \left({s\over s_0}
\right)^{-\lambda_{n,m}}
\beta_m(\Z) \mu_m(\Z_0).
\en
$\beta_0(\Z)$ is the invariant measure of the shape dynamics
($\lambda_{n,0}=0$), and
the functions $\beta_m(\Z)$, $m>0$ are the `excited states' of these dynamics,
describing the decay into the invariant measure, with increasing exponents
$\lambda_{n,m}$. Ref.~\cite{shape} established a correspondence which implies
that the exponents $\lambda_{n,m}$, are precisely the set of
anomalous scaling exponents
of the $n$th order correlation function. The functions $\mu_m$'s are the
adjoint
family of left eigenfunctions of $\gamma_n$, and are the corresponding scaling
structures, namely the zero modes of the Kraichnan operator.

The basic ingredient for the numerical implementation of the shape dynamics for
the calculation of scaling exponents is a finite difference approximation of
the solution of eqs.~(\ref{traject1}) describing the random Lagrangian
trajectories of the $n$ points. We used the simple Ito-Euler method for
this purpose \cite{yellowbook}, which makes the approximation
\eq{traj2}\r_i(t+\delta t)\sim \r_i(t) +\sqrt{\delta t} \eta_i(t),
\en
where the $\eta_i$ are a set of Gaussian random vectors whose covariance as a
function of the $\r_i$'s is given by the spatial part of the covariance of
the $n$ random vectors $\U(\B.r_i)$.

The results of the difference scheme~(\ref{traj2}) are then used to calculate
the instantaneous scale $s(t)$ using some convenient definition of $s$
(we used the square root of the sum of squares of point separations).
We fixed a set of threshold values for $s$ spaced logarithmically, and for
each event
where the value of $s$ increases for the first time beyond one of
the threshold values, we record the instantaneous value of a certain
function of the
shape $\sigma(\Z)$. Once more, the function $\sigma$ is largely arbitrary;
however,
it has to be a permutation-symmetric function of the coordinates.

To extract the relaxation rates, it is convenient to calculate the
autocorrelation of the signal $\sigma(s)$. In fact (for $s<s'$),
\bea \left<\sigma(s)\sigma(s')\right>=\int d\Z d\Z' \beta_0(\Z)
\gamma_n(\Z,\Z',s'/s)\sigma(\Z)\sigma(\Z')=\nonumber\\ \left<\sigma\right>_0^2
+\left<\sigma\mu_1\right>_0\left<\sigma\right>_1
\left({s'\over s}\right)^{-\lambda_{n,1}}
+\cdots,
\ena
where $\left<\right>_m$ denote average with respect to $\beta_m$. Thus,
the autocorrelation function of $\sigma$ relaxes to a constant,
and the asymptotic relaxation rate is
equal to the $n$th order structure function scaling exponent $\zeta_n$.

The precise measurement of the scaling exponent is not easy, however, since
the signal contains a constant whose fluctuation tend to mask the decaying
part which contains the scaling information. Additionally, in the case of
$n=4$ which we consider below, three point contributions contaminate the
signal with relatively persistent transients.
For this reason a very large number of
realizations, $O(10^6)$, is required to measure a scaling exponent, and the
function $\sigma$ needs to be tuned carefully to decrease the effect of
transients\cite{cpu}. Indeed, the autocorrelation for any generic definition
of $\sigma$ decays asymptotically with an exponent $\zeta_n$, however, for
a poorly chosen definition of $\sigma$, the transients may dominate for a
very large range. We have tuned our definitions for $\sigma$ by
trial and error to expose the asymptotic decay rate as early as possible.

We removed the constant from the autocorrelation signal using two methods,
the first was by numerically differentiating the auto-correlation signal,
and the second by subtracting the value of the autocorrelation at the
farthest separation measured ($s/s_0 =100$). Both methods give consistent
results though the numerical differentiation is more noisy while subtraction
tends to introduce a bias in the large $s$ results.

In our $n=3$ calculations we used
\eqs \sigma_3= \log(l_{12})+\log(l_{13})+\log(l_{23})\en and in the four point
\eqs \sigma_4= (l_{12} l_{34})^4+(l_{13} l_{24})^4+(l_{14} l_{23})^4.
\en
 Where $l_{ij}$ is the distance between points $i,j$ normalized such that
$\sum_{i<j}l_{ij}^2 =1$.

The case of third order correlation function exponents deserves attention,
being the
simplest non-trivial one. The Lagrangian paths are relatively easy to
generate in this
case, and the number of realizations needed to observe scaling behavior is
smaller
than in the four point case. Additionally, \cite{3pt-1} demonstrated that three
point correlation
function can be calculated directly by a finite difference scheme, and thus
the two
calculations can be tested against each other for consistency.

We generated a set of data for three point relaxation rates in 4 dimensions for
various values of $\xi$. The number of required realizations was $O(10^5)$,
except for small values of $\xi$ where a larger number is required. We also
generated data for $\xi=1$ for several space dimensionalities, as in the four
point case discussed below. In fig.~1 we display a typical data set measuring
the relaxation of the three point distribution to its invariant measure.
\begin{figure}
\epsfxsize=8truecm
\epsfbox{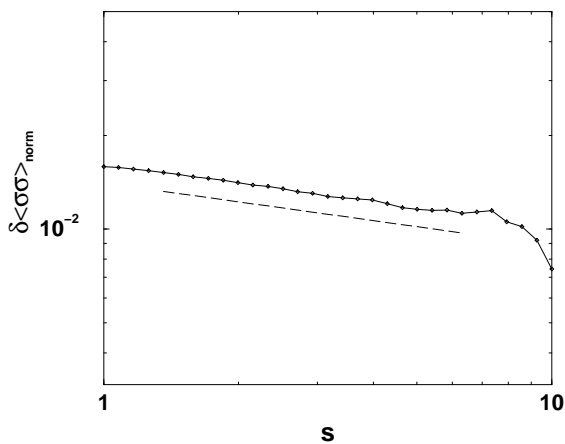}
\caption{The increments of $\left<\sigma_3(s')\sigma_3(s's)\right>$ as a
function
of s, for a three point
simulation in 4 dimensions with $\xi=1$, normalized by the normal scaling
value $s^2$. Taking increments eliminates the constant contribution
$\left<\sigma_3\right>^{2}$. The dashed line is a (shifted) least squares
power law fit. The data are
averaged over $2\times10^5$ realizations}
\end{figure}
Similar results were obtained for the three point simulations for other
 parameter values, with generally a longer scaling range the closer is
$\xi$ to~2.
The results are summarized in fig.~2, along with the leading scaling exponents
calculated by the method of \cite{3pt-1}. We find satisfactory agreement
between
the predictions of the two methods, providing support for the validity of the
present theoretical and numerical framework.

\begin{figure}
\epsfxsize=8truecm
\epsfbox{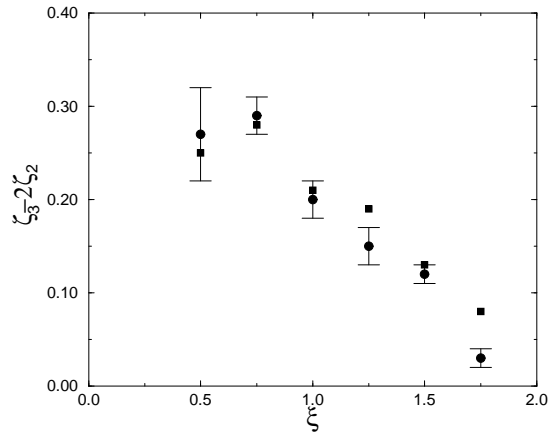}
\caption{The leading three point scaling exponent in four dimensions,
calculated
as a function of $\xi$ by the present Monte-Carlo method (circles) and by the
method of [11]
is difficult to estimate the systematic errors in the results of the latter
method,so no error bars are given. However, we have indications that these
errors
are smaller the statistical errors, except when $\xi$ is near 2. In the
latter case
we use results whose accuracy is higher than those published in[11]}
\end{figure}

Possibly the main interest in our method is in its ability to provide
measurement
of $\zeta_4$ from the
simulation of four point relaxation.
In measuring $\zeta_4$ we concentrated on a scan through various dimensions
and fixed
$\xi=1$. The possibility of performing measurements in various dimensions
(including
the `physical' $d=3$ case) exemplifies the superiority of this algorithm
with respect
to direct numerical simulations, where dimensions larger than 2 are
inaccessible due to memory requirements. The $d$-scan allows us to test,
for the first
time, the perturbative large $d$ expansion results \cite{grisha,shape}.
In fig.~3 we show the results of a typical four point calculation.
\begin{figure}
\epsfxsize=8truecm
\epsfbox{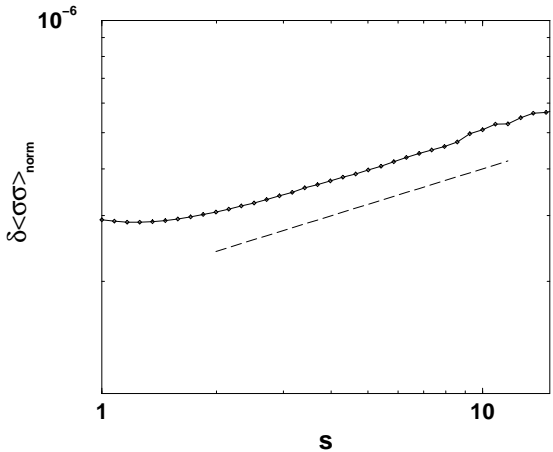}
\caption{The increments of $\left<\sigma_4(s')\sigma_4(s's)\right>$ from a four
point simulation, with the same parameters as in fig~.1, normalized in the same
way, with a power law fit (dashed line). The upward tilt of the graph means
that the
anomaly is positive.
The data are
averaged over $1.6\times10^6$ realizations}
\end{figure}
The results of the three and four point simulations as a function of $d$ are
summarized in fig.~4, along with the prediction of large $d$ perturbation
theory,
and the prediction of Kraichnan's closure. We find that Kraichnan's
prediction agrees
well with the $\zeta_4$ results, whereas the perturbative calculations are
still
not relevant. We cannot estimate how large $d$ should be to agree with the
large $d$
perturbation theory. The prediction for $\zeta_3$ (given in \cite{shape})
is in rough agreement with our simulations.
\begin{figure}
\epsfxsize=8truecm
\epsfbox{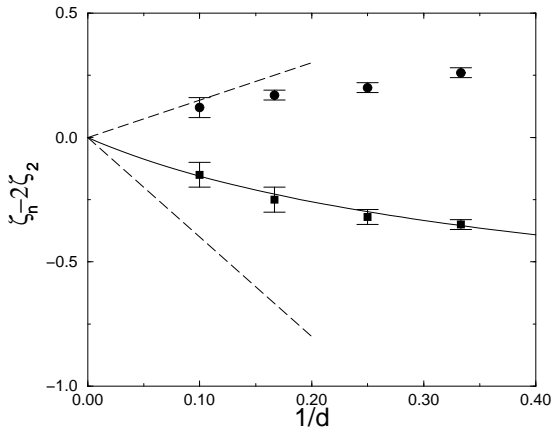}
\caption{The leading three point (circles) and four point (squares) scaling
exponent for velocity scaling exponent $\xi=1$ as a function of the inverse
space dimensionality. The upper and lower dashed lines show the prediction of
first order perturbation theory in $1/d$ for the three and four point cases,
respectively. The continuous line is Kraichnan's closure prediction Eq.(4)}
\end{figure}
In summary, Lagrangian trajectory approach (developed independently by
\cite{massimo})
is seen to provide a numerical tool allowing us to compute the anomalous
exponents of
the Kraichnan model, in principle for the entire parameter space. The
simulations are based
on a new interpretation of anomalous scaling linking it to relaxation rates
in the
stochastic shape
dynamics; the success of the simulations provides further support to this
physically
appealing picture of anomalous scaling in this model. Finally we note that the
basic principles employed here are not specific to the Kraichnan model, and
it might
be possible to use them in other contexts, perhaps sacrificing some of the
simplicity.
In particular, the fact that we need to consider only $n$ points simultaneously
depends heavily on the $\delta$-correlated nature of the velocity, and
would probably
not carry over to more realistic situations.

We thank L. Biferale and U. Frisch for the organization of the the Nice
workshop on
toy models of turbulence where this work was initiated. Part of this work was
done when O. G. and I. P. visited the IHES at Bures-sur-Yvette. We have
benefited from discussions with K. Gawedzki U. Frisch and M. Vergassola. This
work has been supported in part by the European Union under contract
FMRX-CT-96-0010 and the Naftali and Anna Backenroth-Bronicki Fund for
Research in Chaos and Complexity.


\begin{thebibliography}{99}
\bibitem{cpu} We need
approximately 20 cpu-days
to obtain $10^6$ iterations on a typical alpha workstation. Using a cluster
of 8 such
stations a point in parameter space can be calculated relatively accurately
within 2.5 days.
\end{thebibliography}
\end{document}